\documentclass{article}
\usepackage{graphicx} 
\usepackage{hyperref}
\usepackage{authblk}

\title{Privacy-Preserving Large Language Models: Mechanisms, Applications, and Future Directions}
\author[1]{Eric Song}
\author[2]{Guoshenghu Zhao}
\affil[1]{University of California San Diego, email: ejsong@ucsd.edu}
\affil[2]{University of California San Diego, email: g2zhao@ucsd.edu}
\date{November 2024}

\begin{document}

\maketitle












\section*{Abstract}
The rapid advancement of large language models (LLMs) has revolutionized natural language processing, enabling applications in diverse domains such as healthcare, finance and education\cite{vaswani2017attention, brown2020language}. However, the growing reliance on extensive data for training and inference has raised significant privacy concerns, ranging from data leakage to adversarial attacks\cite{xu2021ppml, yan2024llm}. This survey comprehensively explores the landscape of privacy-preserving mechanisms tailored for LLMs, including differential privacy\cite{dwork2006calibrating, abadi2016deep}, federated learning\cite{mcmahan2017communication}, cryptographic protocols\cite{acar2018survey}, and trusted execution environments\cite{xu2021ppml, zhang2024privacy}. We examine their efficacy in addressing key privacy challenges, such as membership inference and model inversion attacks\cite{shokri2017membership, carlini2021extracting}, while balancing trade-offs between privacy and model utility\cite{xu2021ppml, yan2024llm}. Furthermore, we analyze privacy-preserving applications of LLMs in privacy-sensitive domains, highlighting successful implementations and inherent limitations\cite{ullah2023privacypreserving, xiao2024privacymind}. Finally, this survey identifies emerging research directions, emphasizing the need for novel frameworks that integrate privacy by design into the lifecycle of LLMs\cite{xu2021ppml, yang2024ai, hartmann2020privacy}. By synthesizing state-of-the-art approaches and future trends, this paper provides a foundation for developing robust, privacy-preserving large language models that safeguard sensitive information without compromising performance.

\section*{Introduction}
Large Language Models (LLMs), powered by advances in natural language processing, have transformed the way artificial intelligence systems interact with and generate human language. With applications spanning healthcare, finance, education, and beyond, LLMs are increasingly integrated into everyday technologies, offering capabilities such as text generation, summarization, and conversational agents. However, the widespread adoption of LLMs has brought privacy concerns to the forefront. From inadvertent exposure of sensitive information during training to adversarial attacks that exploit model vulnerabilities, ensuring data privacy in LLMs is a critical challenge\cite{yan2024llm, ullah2023privacypreserving, xiao2024privacymind}.
\newline
\newline
LLMs often require extensive datasets for pretraining and fine-tuning, which may include sensitive, proprietary, or personally identifiable information (PII). These datasets, coupled with the ability of the models to memorize and reproduce training data, expose risks such as data leakage, membership inference, and model inversion attacks\cite{xu2021ppml, zhang2024privacy, xiao2024privacymind, chaudhuri2011differentially, jordon2018pate}. Moreover, privacy-preserving measures can significantly impact the utility and efficiency of LLMs, presenting a delicate trade-off between safeguarding sensitive data and maintaining model performance.
\newline
\newline
To address these challenges, researchers have developed a range of privacy-preserving mechanisms tailored to LLMs. Techniques such as differential privacy, federated learning, cryptographic protocols, and trusted execution environments have demonstrated varying degrees of success in mitigating privacy risks while retaining usability\cite{qin2024local, xu2021ppml, zhang2024privacy}. These approaches are particularly crucial in privacy-sensitive domains, including healthcare and finance, where breaches of data confidentiality can have severe consequences\cite{ullah2023privacypreserving, yang2024ai}.
\newline
\newline
This paper provides a comprehensive survey of privacy-preserving mechanisms for LLMs, exploring their technical foundations, real-world applications, and limitations\cite{hartmann2020privacy, yan2024llm}. We categorize existing methods, evaluate their effectiveness against specific privacy threats, and discuss the trade-offs they entail. Furthermore, we examine applications of LLMs in scenarios requiring stringent privacy guarantees and highlight case studies demonstrating the integration of privacy by design principles. Finally, we identify emerging trends and research directions, emphasizing the need for frameworks that seamlessly embed privacy-preserving features into the lifecycle of LLMs\cite{xu2021ppml, zhang2024privacy, xiao2024privacymind}.
\newline
\newline
By synthesizing state-of-the-art methods and future trajectories, this paper aims to equip researchers and practitioners with insights and tools to develop LLMs that uphold data privacy without compromising functionality. In doing so, it addresses the growing imperative for privacy-preserving AI in an era of pervasive data-driven technologies.

\section*{Privacy Challenges in Large Language Models}

LLMs pose significant privacy challenges due to their reliance on extensive datasets and their generative capabilities. One of the primary risks is the inadvertent exposure of sensitive information during training and inference. This is particularly concerning in domains like healthcare and finance, where sensitive data is often used. Models trained on such data may unintentionally memorize specific details, which can be extracted during inference, either inadvertently or through deliberate attempts by malicious actors.
\newline
\newline
Membership inference attacks represent another critical threat to LLMs. These attacks exploit the model's responses to determine whether a specific data point was part of the training set. Such vulnerabilities not only expose sensitive or proprietary information but also raise concerns about compliance with privacy regulations, such as GDPR and HIPAA. Beyond this, model inversion attacks add another layer of risk by enabling adversaries to reconstruct sensitive input data based on the model’s outputs.
\newline
\newline
The challenges are further compounded by the trade-offs between implementing effective privacy-preserving mechanisms and maintaining model utility. Techniques like differential privacy or federated learning often degrade model performance, leading to a delicate balance between safeguarding data privacy and preserving the usability and efficiency of the model. These issues highlight the urgent need for robust, scalable solutions that integrate privacy-preserving principles throughout the lifecycle of LLMs. Addressing these challenges is critical for the safe and responsible deployment of LLMs in privacy-sensitive applications.

\section*{Mechanisms for Privacy Preservation}
\subsection*{Differential Privacy}
Differential Privacy (DP) is a robust mathematical framework for safeguarding individual data points during training and inference. An algorithm \( M \) satisfies \(\epsilon\)-Differential Privacy if, for any two datasets \( D \) and \( D' \) differing by a single element, and any output \( O \), the following holds:
\[
\frac{\Pr[M(D) = O]}{\Pr[M(D') = O]} \leq e^{\epsilon},
\]
where \(\epsilon > 0\) quantifies the privacy-utility trade-off, with smaller values offering stronger privacy guarantees.
\newline
\newline
In training large language models (LLMs), DP is often implemented via differentially private stochastic gradient descent (DP-SGD). Gradients are clipped to a threshold \( C \) to limit individual influence:
\[
g_i' = g_i \cdot \min\left(1, \frac{C}{\|g_i\|}\right),
\]
and Gaussian noise \( N(0, \sigma^2) \) is added to the aggregated gradients:
\[
\tilde{g} = \frac{1}{n} \sum_{i=1}^n g_i' + N(0, \sigma^2).
\]
This ensures privacy at the cost of model utility, with the trade-off determined by the noise scale \(\sigma\) and privacy budget \(\epsilon\).
\newline
\newline
Despite its effectiveness, DP faces challenges in LLMs, such as computational overhead from gradient clipping and noise addition, and performance degradation in tasks requiring high precision. Additionally, applying DP during inference, such as through output perturbation, risks reducing coherence in generated text.
\newline
\newline
To mitigate these issues, researchers are exploring hybrid methods combining DP with cryptographic techniques or federated learning. Adaptive privacy budgets, which adjust \(\epsilon\) based on task requirements, also show promise in balancing privacy and utility.
\newline
\newline
By integrating DP into the LLM lifecycle, robust protections against data leakage can be achieved, though challenges in scalability and utility remain areas for further research.

\subsection*{Federated Learning}
Federated Learning (FL) is a decentralized training approach enabling large language models (LLMs) to learn collaboratively across distributed data sources while keeping sensitive data localized on devices. Instead of sharing raw data, FL relies on exchanging model updates, such as gradients or weights, which are aggregated to form a global model, reducing privacy risks.
\newline
\newline
In FL, the global model is updated iteratively as follows:
\[
w^{t+1} = w^t + \eta \sum_{i=1}^n \frac{n_i}{N} \Delta w_i^t,
\]
where \( w^t \) is the global model at round \( t \), \( \Delta w_i^t \) is the update from client \( i \), \( n_i \) is the size of client \( i \)'s dataset, \( N \) is the total dataset size, and \( \eta \) is the learning rate.
\newline
\newline
FL offers strong privacy benefits by ensuring that data remains on devices, making it well-suited for privacy-sensitive applications such as healthcare and personalized AI. However, challenges persist, including handling non-IID (non-independent and identically distributed) data and mitigating privacy risks from shared gradients. To address these, techniques like differential privacy are employed, where Gaussian noise is added to gradients:
\[
\tilde{g}_i = g_i + N(0, \sigma^2),
\]
where \( g_i \) is the original gradient and \( N(0, \sigma^2) \) is noise with standard deviation \( \sigma \).
\newline
\newline
Despite these advantages, FL faces communication overheads when scaling to LLMs with billions of parameters. Methods like gradient sparsification and quantization are often used to reduce bandwidth usage. By integrating FL into LLM training, organizations can achieve a balance between privacy and performance, though challenges in scalability and efficiency remain active research areas.

\subsection*{Cryptographic Methods}
Cryptographic methods ensure robust privacy preservation by encrypting data during the training and inference of large language models (LLMs). Unlike techniques like differential privacy, cryptographic approaches secure data even in untrusted environments, making them ideal for collaborative training and applications requiring strict confidentiality.
\newline
\newline
\textbf{Homomorphic encryption (HE)} enables computations directly on encrypted data. Formally, for plaintext inputs \( x_1 \) and \( x_2 \), HE satisfies the following:
\[
Dec(Enc(x_1) \circ Enc(x_2)) = x_1 \ast x_2,
\]
where \( Enc \) and \( Dec \) are the encryption and decryption functions, and \( \circ \) represents operations on ciphertexts. While HE provides strong privacy guarantees, its high computational overhead limits its scalability to large LLMs.
\newline
\newline
\textbf{Secure multi-party computation (SMPC)} allows multiple parties to compute a function \( f(x_1, x_2, \ldots, x_n) = y \) over their inputs \( x_i \) without revealing them. Data is split into secret shares, ensuring privacy during collaborative training, though communication overhead can become a bottleneck in large-scale systems.
\newline
\newline
\textbf{Trusted execution environments (TEEs)} provide a hardware-based approach by creating secure execution zones within processors. TEEs enable the processing of sensitive data during training or inference, ensuring data confidentiality and integrity without exposing it to external entities.
\newline
\newline
Although cryptographic methods offer strong privacy guarantees, challenges such as high computational costs for HE and communication overhead for SMPC hinder their scalability for LLMs. However, these methods have shown promise in privacy-sensitive domains like healthcare and finance. Advances in hybrid approaches and hardware acceleration are making cryptographic methods increasingly practical for real-world LLM applications.
\newline
\newline
By incorporating cryptographic techniques into LLM workflows, organizations can achieve strong privacy protection, though addressing efficiency and scalability remains a critical area for research.



\section*{Applications of Privacy-Preserving LLMs}
Privacy-preserving large language models (LLMs) have transformative potential across various domains where sensitive data is involved. By integrating privacy mechanisms like differential privacy, federated learning, and cryptographic methods, LLMs can be deployed in high-stakes applications while ensuring data confidentiality and regulatory compliance.
\newline
\newline
\textbf{Healthcare}: In healthcare, privacy-preserving LLMs are revolutionizing how sensitive patient data is utilized. These models can assist in medical diagnosis, drug discovery, and personalized treatment recommendations by analyzing distributed electronic health records (EHRs) without centralizing sensitive information. For instance, federated learning enables multiple hospitals to collaboratively train LLMs on their localized datasets, mitigating the risk of data breaches. Privacy techniques like differential privacy further ensure that individual patient records remain confidential, even during model inference. Such deployments are crucial for complying with regulations like HIPAA and GDPR while enabling advancements in AI-powered medical research.
\newline
\newline
\textbf{Finance}: The financial sector often handles proprietary and sensitive information, making it a prime candidate for privacy-preserving LLMs. These models can be used for fraud detection, risk assessment, and anti-money laundering analysis while maintaining strict data confidentiality. For example, cryptographic methods like secure multi-party computation allow institutions to jointly analyze transaction data without sharing raw datasets. Homomorphic encryption is particularly useful for processing encrypted financial data directly, enabling secure customer profiling or credit scoring. By protecting individual and institutional data, LLMs also support compliance with financial privacy laws such as the CCPA and PSD2.
\newline
\newline
\textbf{Education and Public Services}: In education and public service, privacy-preserving LLMs facilitate personalized learning, citizen engagement, and efficient service delivery while safeguarding sensitive information. In education, LLMs can analyze student performance data to generate adaptive learning plans without exposing individual identities. Similarly, in public services, these models can process demographic data to optimize resource allocation or provide personalized assistance, such as in tax filing or social benefit applications, without risking privacy breaches. Federated learning and trusted execution environments are often deployed to maintain security and efficiency in such contexts.
\newline
\newline
\textbf{Emerging Scenarios}: Privacy-preserving LLMs are increasingly finding applications in emerging areas like smart cities, autonomous systems, and cross-border data collaborations. In smart cities, LLMs can analyze distributed IoT data to optimize traffic management, energy distribution, and public safety while ensuring data confidentiality. In autonomous systems, such as self-driving vehicles, privacy-preserving models process sensitive environmental and user data without exposing critical information to external entities. Furthermore, cross-border collaborations in research and policymaking rely on secure data sharing and processing, where cryptographic methods like homomorphic encryption and differential privacy play vital roles.

\section*{Conclusion}
This paper provides a comprehensive exploration of privacy-preserving mechanisms for large language models (LLMs), addressing critical challenges such as data leakage, membership inference, and model inversion attacks. By evaluating approaches like differential privacy, federated learning and cryptographic protocols, we highlight their effectiveness and the trade-offs between privacy and model utility. Our work underscores the transformative potential of privacy-preserving LLMs in sensitive domains such as healthcare, finance, and education, while identifying limitations and future research opportunities. By advocating for privacy-by-design principles and synthesizing state-of-the-art techniques, this paper lays the groundwork for advancing secure, responsible, and high-performing LLMs in an increasingly data-driven world.

\bibliographystyle{plain}
\bibliography{references}


\end{document}